\documentclass[graybox]{svmult}

\usepackage{mathptmx}       
\usepackage{helvet}         
\usepackage{courier}        
\usepackage{type1cm}        
\usepackage{makeidx}         
\usepackage{graphicx}        
\usepackage{multicol}        
\usepackage[bottom]{footmisc}

\makeindex             

\begin{document}

\title*{Application of the Bead Perturbation
Technique to a Study of a Tunable 5 GHz Annular Cavity}
\author{Nicholas M Rapidis}
\institute{Nicholas M Rapidis \at University of California Berkeley, Berkeley, CA 94720, \email{rapidis@berkeley.edu}}
%
%
\maketitle

\abstract*{Microwave cavities for a Sikivie-type axion search are subject to several constraints. In the fabrication and operation of such cavities, often used at frequencies where the resonator is highly overmoded, it is important to be able to reliably identify several properties of the cavity. Those include identifying the symmetry of the mode of interest, confirming its form factor, and determining the frequency ranges where mode crossings with intruder levels cause unacceptable admixture, thus leading to the loss of purity of the mode of interest. A simple and powerful diagnostic for mapping out the electric field of a cavity is the bead perturbation technique. While a standard tool in accelerator physics, we have, for the first time, applied this technique to cavities used in the axion search. We report initial results from an extensive study for the initial cavity used in the HAYSTAC experiment. Two effects have been investigated: the role of rod misalignment in mode localization, and mode-mixing at avoided crossings of TM/TE modes. Future work will extend these results by incorporating precision metrology and high-fidelity simulations.}

\abstract{Microwave cavities for a Sikivie-type axion search are subject to several constraints. In the fabrication and operation of such cavities, often used at frequencies where the resonator is highly overmoded, it is important to be able to reliably identify several properties of the cavity. Those include identifying the symmetry of the mode of interest, confirming its form factor, and determining the frequency ranges where mode crossings with intruder levels cause unacceptable admixture, thus leading to the loss of purity of the mode of interest. A simple and powerful diagnostic for mapping out the electric field of a cavity is the bead perturbation technique. While a standard tool in accelerator physics, we have, for the first time, applied this technique to cavities used in the axion search. We report initial results from an extensive study for the initial cavity used in the HAYSTAC experiment. Two effects have been investigated: the role of rod misalignment in mode localization, and mode-mixing at avoided crossings of TM/TE modes. Future work will extend these results by incorporating precision metrology and high-fidelity simulations.}

\section{Background}
\label{sec:1}

\subsection{Introduction}
\label{subsec:2}
 HAYSTAC is a University of California, Berkeley, University of Colorado Boulder, and Yale University experimental collaboration designed to detect the QCD axion. The experiment consists of a Sikivie type detector that detects the axion through its conversion into two photons via the Primakoff effect. One photon is virtual and is provided by a strong magnetic field \cite{Sikivie}. The ultimate range of the experiment would allow for the detection of axion masses in the 20--100 $\mu eV$ range. A 9.4 T magnet provides the required magnetic field. A VeriCold Technologies dilution refrigerator maintains the physical temperature of the microwave cavity and the Josephson Parametric Amplifier at 127 mK. The microwave cavity setup consists of a 10 in. height and 4 in. diameter oxygen-free high conductivity copper cavity. A 2 in. diameter copper rod is used to tune the cavity's $\mathrm{TM}_{010}$-like mode. The axis of the tuning rod is off-center from the axis of the cavity; when the rod is rotated $180^\circ$, the frequency  of the $\mathrm{TM}_{010}$-like mode tunes over 3.6--5.8 GHz. \cite{NIM}.  

\subsection{Electromagnetic Properties of the Resonator}
\label{subsec:2}

The axion conversion power in a microwave cavity is given by \cite{NIM}:
\begin{equation}
P_{sig} \propto B^2_0 V C_{nml} Q_L. 
\end{equation}

For optimal performance, it is therefore ideal to maximize the form factor ($C_{nml}$) and the quality factor ($Q_L$). $Q_L$ is the quality factor of a resonator which is proportional to a mode-dependent constant of order 1 and the volume and inversely proportional to the surface area and the skin depth. The form factor is defined as: 

\begin{equation}
C_{nml} \equiv \frac{(\int d^3 \textbf{x} \: \mathbf{\hat{z}} \cdot \mathbf{e^{\ast}_{nml}} (\mathbf{x}))^2}{V\int d^3 \textbf{x} \: \epsilon (\mathbf{x}) \: |{\mathbf{e_{nml}} (\mathbf{x})}|^2}
\end{equation}

where V is the volume of the vacuum inside the cavity and $\epsilon(\textbf{x})$ is the dielectric constant, which is set to 1.
The $\mathrm{TM_{010}}$-like mode of the cavity is used because it posseses a uniform electric field along the z-direction which gives the largest form factor of any mode.

The resonant frequencies of TM-like modes are a function of rod position whereas TE and TEM modes show virtually no dependence (Figure 1). This allows for a wide range of potential axion masses to be scanned by tuning the TM modes. However, problems are introduced due to mode mixing when the frequency of the desired TM mode approaches the frequency of a TE or TEM mode. 

When the frequency of the desired TM mode approaches a stationary mode (a TE or TEM mode) mode mixing occurs. The two modes in proximity hybridize, i.e. they become linear combinations of the two states as they are far apart, and thus the form factor of the initial TM-like mode diminishes by an unknown amount. The phenomenon is analogous to a two-level avoided crossing in quantum mechanics, seen in atomic and nuclear systems. This leads to regions in the frequency spectrum where the sensitivity of the experiment to axions is compromised, and thus needs to be excised from the data set.  What is more problematic is that each mode crossing is different, and it is unknown over what frequency span, the data should be cut out.  The bead perturbation technique allows us to examine the hybridization of the modes as they approach one another, and thus guide the decision of what frequency range should be eliminated.  Optimistically, it could allow us to recover some or most of the data in the mode-mixing region by measuring the TM component of the two mixed states. The results reported here are directed towards the study of such phenomena.

%
\begin{figure}[t]
\sidecaption
\includegraphics[scale=.40]{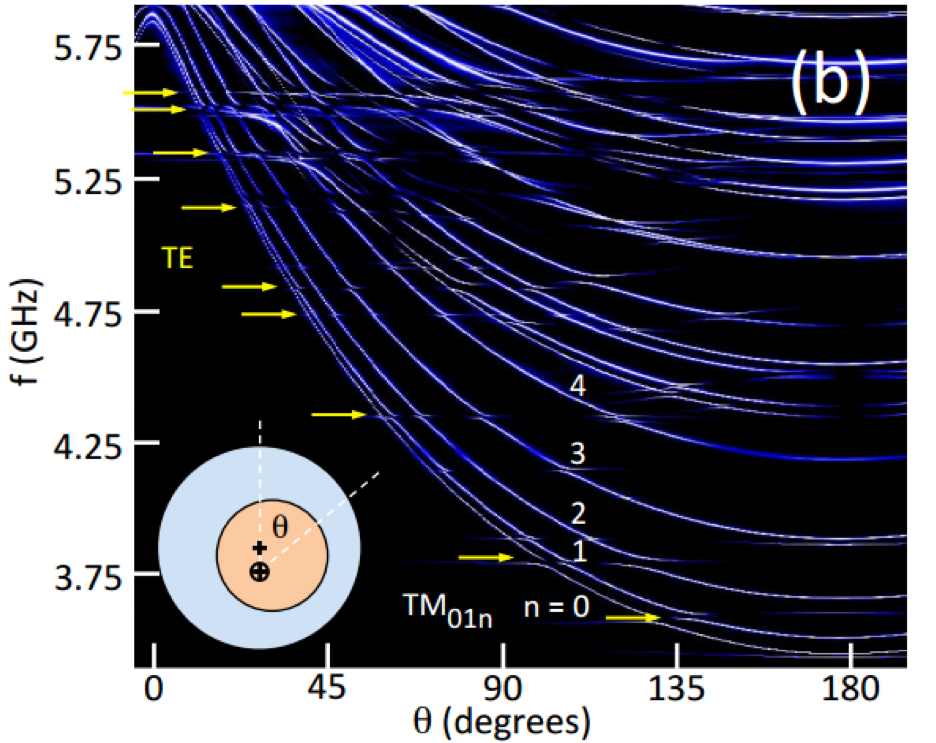}
%
%
\caption{Mode frequencies as a function of tuning rod angle. The TM modes are tuned through the TE modes whose frequencies are only weakly dependent on rod angle.}
\label{fig:1}       
\end{figure}

\subsection{Bead Perturbation Technique}
\label{subsec:2}

A useful tool in observing properties of certain cavity modes is the bead perturbation technique. In test cavities, a hole is made on both endcaps such that a small dielectric bead can pass through the whole length of the cavity at a constant radial and angular position. In our studies, a cylindrically shaped alumina ($\epsilon = 9.1$) bead of $h=4.80 \, \mathrm{mm}$ and $r=2.15 \, \mathrm{mm}$ attached to a Kevlar string traverses the cavity. As the bead travels through the cavity, the resonant frequencies of the cavity modes undergo small frequency shifts that can be calculated in perturbation theory. Specifically:

\begin{equation}
\frac{\Delta \omega}{\omega} = \frac{-(\epsilon - 1)}{2} \frac{V_{Bead}}{V_{Cavity}} \frac{E(\textbf{r})^2}{\langle E(\textbf{r})^2 \rangle_{cav}} 
\end{equation}

where V is the volume and $\epsilon$ is the dielectric constant of the bead \cite{JCS}.

This method provides a ``profile'' of each mode at a given rod angle by plotting the frequency of the mode as a function of bead position inside the cavity. While the frequency shift only determines the magnitude of the electric field and not its individual vector components, comparing the profile with a simulation invariably allows characteristics of modes to be determined.

\section{Mode Mixing}
\label{sec:2}

Mode mixing can cause significant regions of the frequency range to become unusable due to a potentially lower $Q_L$ and a lower form factor of the $\mathrm{TM}_{010}$-like mode. To study the effect of these mode crossings, the $\mathrm{TM}_{010}$-like mode was tuned to a frequency with no noticeable mixing. This was confirmed by performing a bead pull and confirming that both the $\mathrm{TM}_{010}$-like mode and the intruder mode (TE or TEM mode) that were being studied, had the expected profiles. The $\mathrm{TM}_{010}$-like mode was almost entirely flat whereas the intruder modes had several peaks and nodes. The $\mathrm{TM}_{010}$-like mode was tuned such that its frequency increased and approached the frequency of the intruder mode. This was done by adjusting the angle of rotation of the tuning rod in small increments. As the two modes approached one another they mixed, as is evident from the field profiles in the insets of Figure 2.  The lower frequency $\mathrm{TM}_{010}$-like mode developed an oscillatory profile, whereas the higher frequency TE mode picked up a constant offset component.  In this case (though not in all such cases), the two hybrid modes never crossed; as the $\mathrm{TM}_{010}$-like mode frequency continued to be changed, the lower mode remained stationary and asymptotically became the TE mode, whereas the higher mode moved upward in frequency and asymptotically became the pure $\mathrm{TM}_{010}$-like mode. The mixing could also be observed by recording the $Q_L$ of the $\mathrm{TM}_{010}$-like mode through a vector network analyzer. The $\mathrm{TM}_{010}$-like mode was then tuned to a frequency higher than that of the intruder mode until it was no longer exhibiting any mixing, thus determining the range of the frequencies over which mode mixing occurs. Throughout this process, a field profile was measured for both the lower and upper mode.

\begin{figure}[ht]
\sidecaption
\includegraphics[scale=.40]{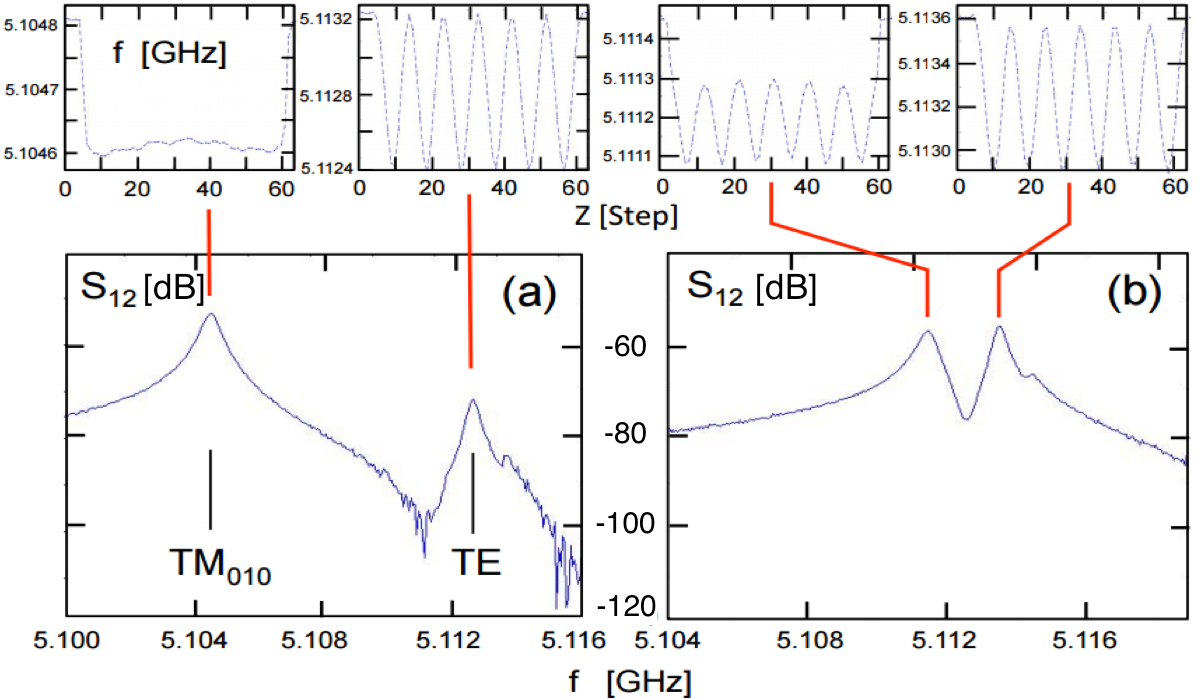}
%
%
\caption{Bead pull profiles for the frequency of the $\mathrm{TM}_{010}$-like mode and a TE mode. (a) corresponds to a rod position where the two peaks are not mixing. In (b) the lower peak ($\mathrm{TM}_{010}$-like) exhibits an oscillatory behavior as a function of the position of the bead, in contrast to the case where no mixing occurrs where such variations are effectively absent. The cavity extent is approximately $z = 5$ to $z = 60$ \cite{NIM}.}
\label{fig:2}       
\end{figure}

While this investigation on mode crossing behavior did confirm the expected hybridization of the $\mathrm{TM}_{010}$-like mode as it crossed intruder modes, a lack of mixing was observed for several intruder modes. Many modes in the frequency range of the $\mathrm{TM}_{010}$-like mode were benign even though they were noticeable peaks in the spectrum. 

\section{Grid Measurements}
\label{sec:3}

The profile of modes as observed through the bead perturbation technique is highly sensitive to any breaking of the axial symmetry in a cylindrical cavity. If the axis of the rod is not parallel to the axis of the cavity, the $E_z$ component of the field is no longer constant along the axis. Instead, there is localization which results in a difference in the magnitude of the frequency shift introduced by the bead at different z positions. It is therefore worth studying the effect of small misalignments in the rod's tilt.

To control misalignments, two orthogonally oriented micrometers were placed on one of the endcaps. The micrometers were positioned around the axel of the copper rod, allowing one end of the rod's axis to be translated perpendicularly to the axis of the cavity. This corresponded to a tilt in the axis of the rod with respect to the axis of the cavity. The rod was locked at a specific angle by placing a tight collar around its axel, thus inhibiting rotations. 

By performing a bead pull, the resulting frequency shift due to tilting was observed. At each misalignment position, a change in frequency ($\Delta f$) was recorded which corresponded to the frequency difference of the $\mathrm{TM}_{010}$-like mode with the bead at one end of the cavity compared to the other end (Figure 3).

\begin{figure}[ht]
\sidecaption
\includegraphics[scale=.38 ]{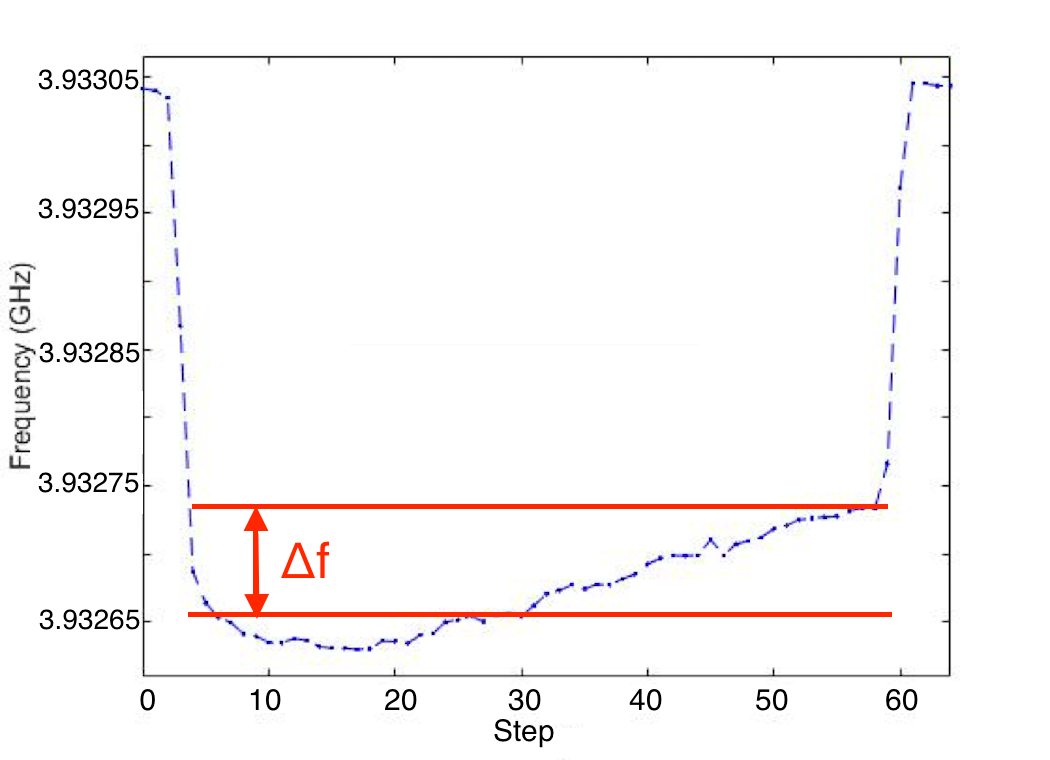}
%
%
\caption{Non-flat profile of the mode frequency for a bead pull when symmetry was broken. The cavity extent is from z $=$ 5 to z $=$ 60. In this case $\Delta f \equiv f(z = 5) - f(z = 60)$}
\label{fig:3}       
\end{figure}

The micrometers were adjusted to take a grid of points around the central position where $\Delta f = 0$. This procedure was executed in three different tuning positions: $\theta = 0^\circ$ corresponding to the rod being at the center of the cavity, $\theta = 90^\circ $, and $\theta = 180^\circ $.

Similar behavior was observed in all three tuning positions. The data formed a plane-like surface with a series of nearly flat points located on one of the diagonals (Figure 4). Computing the tilt of the rod's axis with respect to the cavity's, one can calculate that each step of 0.8 mil (0.0008 in.) displacement corresponds to an angle difference of about 0.08 mrad. It is thus clear that the cavity is very sensitive to rod misalignments and hence the rod alignment needs to be carefully controlled. 

\begin{figure}[ht]
\sidecaption
\includegraphics[scale=.24]{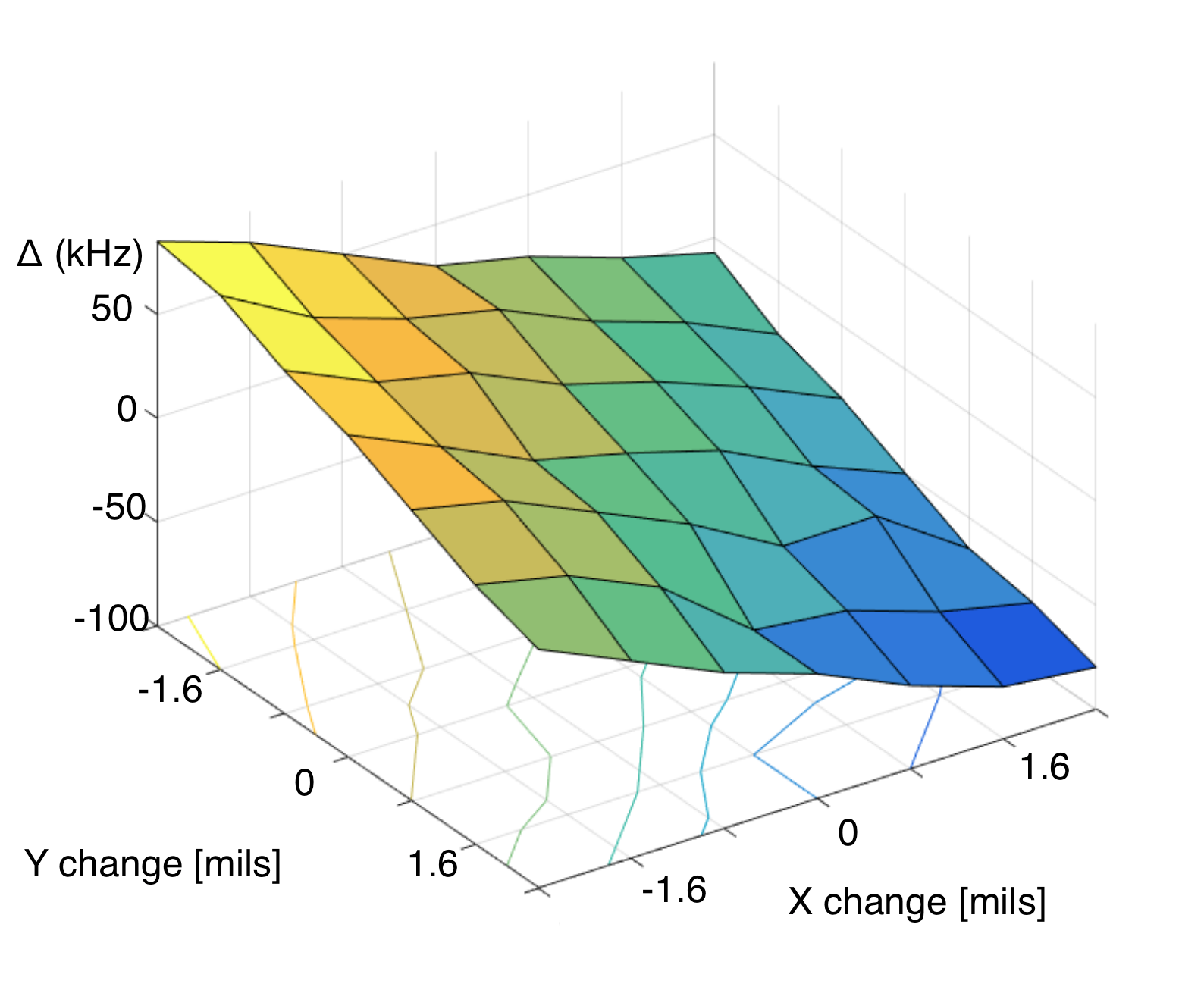}
%
%
\caption{$\Delta f$ as a function of micrometer displacements in the $\theta = 90^\circ$ case.}
\label{fig:2}       
\end{figure}

\section{Conclusion and Future Work}
\label{sec:4}

A better understanding of the effects of mode crossings and misalignments in the HAYSTAC microwave cavity has been achieved leading to a better understanding of potential causes of a loss of the signal power. The effects of small rod misalignments are significant since they establish the dimensional and angular tolerances for cavity fabrication and assembly. 

Future work on these aspects of the experiment will include a comparison with simulations on CST Microwave Studio. These simulations will  have an accurate rendering of the rod used in this investigation created from highly precise metrology. An \textit{in situ} bead perturbation capability is being contemplated for the actual HAYSTAC experiment for real time monitoring of the quality of the mode from which data is being measured.  An open question is whether there are any misalignments induced when the cavity is cooled to $<$100 mK.

A forthcoming paper will report on this work and on simulations on the measurements.

\begin{acknowledgement}
This work was supported under the auspices of the National Science Foundation, under grant PHY-1607417, and the Heising-Simons Foundation under grant 2014-182.
\end{acknowledgement}
%

%
%

\begin{thebibliography}{99.}%
%
%
%

\bibitem{Sikivie} P. Sikivie, Phys. Rev. D \textbf{32}, 2988 (1985). doi:10.1103/PhysRevD.36.974
%
\bibitem{NIM} S. Al Kenany \textit{et. al.}, Nucl. Inst. Meth. Phys. Res. A \textbf{854}  (2017), 11. doi:10.1016/j.nima.2017.02.012
%
\bibitem{JCS} J. C. Slater, Rev. Mod. Phys. \textbf{18} (1946) 441. doi:10.1103/RevModPhys.18.441.




\end{thebibliography}
%

\end{document}